\documentclass[twocolumn,english,prb,showpacs]{revtex4-1}
\usepackage[latin9]{inputenc}
\usepackage{amstext,bm}
\usepackage{graphicx}
\usepackage{xcolor}
\usepackage{amsmath}
\usepackage{babel}
\usepackage[titletoc]{appendix}
\usepackage{hyperref}
\usepackage{float}

\begin{document}

\title{{Node-Surface and Node-Line Fermions From Nonsymmorphic Lattice Symmetries}}

\author{Qi-Feng Liang$^{1}$}\email{qfliang@usx.edu.cn} \author{Jian Zhou$^2$} \author{Rui Yu$^3$} \author{ Zhi Wang$^{4}$} \author{Hongming Weng$^{5, 6}$}\email{hmweng@iphy.ac.cn}

\address{$^{1}$ Department of Physics, Shaoxing University, Shaoxing 312000, China}
\address{$^{2}$ Department of Materials, Nanjing University, Nanjing, China}
\address{$^{3}$ Department of Physics, Harbin Institute of Technology, Harbin}
\address{$^{4}$ School of Physics and Engineering, Sun Yat-sen University, Guangzhou 510275, China}
\address{$^{5}$ Beijing National Laboratory for Condensed Matter Physics, and Institute of Physics, Chinese Academy of Sciences, Beijing 100190, China}
\address{$^{6}$ Collaborative Innovation Center of Quantum Matter, Beijing 100190, China}
\begin{abstract}
We propose a kind of novel topological quantum state of semimetals in a quasi-one-dimensional (1D) crystals BaMX$_3$ (M = V, Nb or Ta; X = S or Se) family by using symmetry analysis and first principles calculation. We find that in BaVS$_3$ the valence and conduction bands are degenerate in the $k_z=\pi/c$ plane ($c$ is the lattice constant along $\hat{z}$ axis) of the Brillouin Zone (BZ). These nodal points form a node-surface and they are protected by a nonsymmorphic crystal symmetry consisting of a two-fold rotation about the $\hat{z}$ axis and a half-translation along the same $\hat{z}$ axis. The band degeneracy in the node-surface is lifted in BaTaS$_3$ by including strong spin-orbit coupling (SOC) of Ta. The node-surface is reduced into 1D node-lines along the high-symmetry paths $k_x=0$ and $k_x$ = $\pm{\sqrt{3}}k_y$ on the $k_z=\pi/c$ plane. These node-lines are robust against SOC and guaranteed by the symmetries of $P6_3/mmc$ space group. These node-line states are entirely different from previous proposals which are based on the accidental band touchings. We also propose a useful material design for realizing topological node-surface and node-line semimetals.
\end{abstract}
\maketitle

\section{Introduction}
Searching for new topological states of matters becomes an active field since the discovery of topological insulators (TI)\cite{Hasan_Kane_RMP_2010,Qi_Zhang_RMP_2011}. Recently,
much attention has been draw to the topological semimetal (TSM) due to the stimulation of successful design\cite{Weng_PRX2015} and experimental observations of Weyl semimetal 
(WSM) in transition-metal monophosphides \cite{Xu_2015,LvBQ_PRX2015}. WSM
are topological metallic states in which the Fermi surfaces are consisted of discrete two-fold-degenerate Weyl points. These Weyl points are topologically robust since 
they carry nontrivial chiral charges. The idea of TSMs has also been extended to node-line semimetal (NLS) hosting one-dimensional (1D) contacts of conduction 
and valence bands.~\cite{burkov} The node-line fermions have been proposed in realistic materials such as Graphene-network\cite{WengHM_PRB2015}, 
Cu$_3$N\cite{Kim_PRL2015}, Cu$_3$PdN\cite{YuRui_PRL2015} and Ca$_3$P$_2$\cite{XieCava_APLM2015}. Experimental characterization of 
topological properties of NLSs was carried out for PbTaSe$_2$\cite{BianG_arxiv2015}. {One of the important topological properties of NLSs is that they support ``drumhead" like flat 
surface band~\cite{burkov, WengHM_PRB2015} which may potentially enhance the superconductivity transition temperature\cite{Heikkila_arxiv2015}}.

In previous proposals,~\cite{WengHM_PRB2015,Kim_PRL2015,YuRui_PRL2015} ring-like node-line contacts are fragile against spin-orbit coupling (SOC). Including SOC lifts 
the node-line contacts and drives the systems into topological insulators or WSM. Therefore, searching for stable NLSs robust to SOC~\cite{FangC_PRB2015} is of great interest 
and importance. Recently, materials design principles involving a layer-stacking process of topological insulators and magnetic insulators are given for WSM~\cite{Burkov_PRL2011} 
and NLSs\cite{Phillips_PRB2014}. Alternatively, in the present work we provide a new route to realize such exotic TSM state by arranging one-dimensional ionic chains in parallel. 
By using first principles calculations and symmetry consideration, we show that the quasi-one-dimensional crystal BaVS$_3$ exhibits a two-dimensional (2D) touching of valence and 
conduction bands, namely a node-surface, at the $k_z=\pi/c$ plane with $c$ being the $\hat{z}$-axis lattice constant. In a cousin compound BaTaS$_3$, the large SOC of Ta atom  lifts 
the 2D degeneracy and reduces it into 1D node-lines which are robust to SOC. The node-lines found in the present work locate at the high-symmetry paths $k_x=0$ and 
$k_x=\pm{\sqrt{3}}k_y$ on the $k_z=\pi/c$ plane, demonstrating their purely symmetrical origin that is different from the accidental-band-degeneracy mechanism of previous 
node-lines\cite{BianG_arxiv2015,FangC_PRB2015}.
\section{Crystal structure}

\begin{figure}[t]
\centering{}
\includegraphics[width=0.5\textwidth]{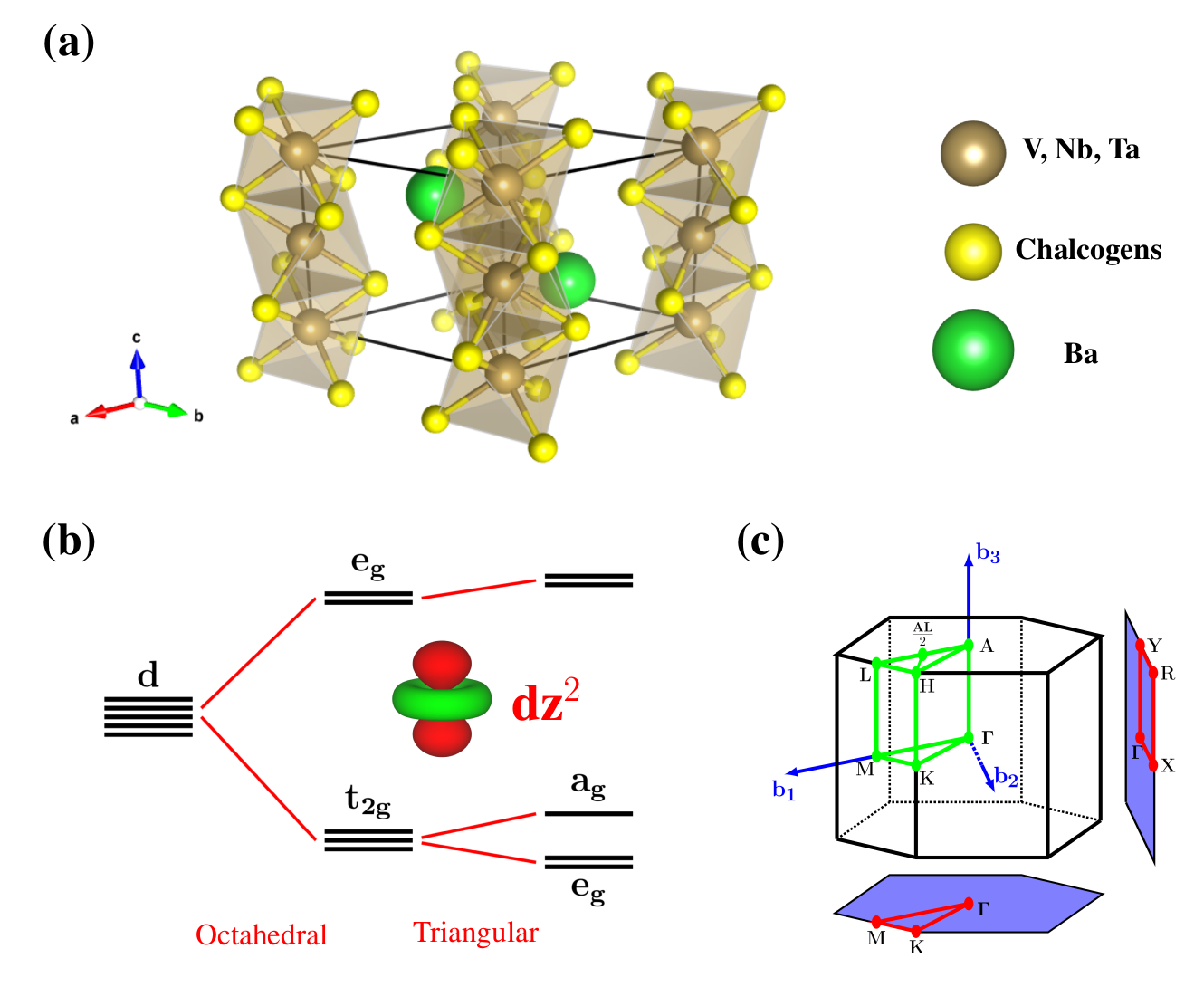}
\caption{ (Color online) (a) Unit cell of Hexagonal BaMX$_3$. Large (green), mediate (yellow) and small (brown) spheres denote the Ba, M and X atoms. (b) Splitting of $d$-orbitals under the crystal fields. The crystal field of MX$_3$-octahedron splits the five-fold $d$ orbitals into two-fold $e_g$ and three-fold $t_{2g}$ orbitals. The crystal-field of the hexagonal lattice further splits the three-folded $t_{2g}$ orbitals into two-folded $e'_g$  and $a_g$ orbitals. $a_g$ has the character of d$_{3z^2-r^2}$ orbital shown in the inset. (c) The Brillouin Zone of BaMX$_3$. Shadowed planes are the projected 2D Brillouin zone to different crystal facets.
\label{fig:crystal} }
\end{figure}

The BaMX$_3$ (M = V, Nb or Ta; X = S or Se) is a group of quasi-one-dimensional crystals adopting hexagonal structure with the space group of \emph{P}6$_3$/\emph{mmc} (No. 194)\cite{Gardner_BVS1969,Gardner_BTaS1969} as shown in Fig.~\ref{fig:crystal}. M atoms are surrounded by octahedron of X atoms and these octahedrons  form linear chains along the $\hat{z}$-axis by sharing common surfaces. Those chains are lined up and arranged into a trigonal 
lattice in the $x-y$ plane with Ba atoms filling the space between the chains. As the inter-chain distance is much larger than the intra-chain distance, these materials are considered as quasi-one-dimensional crystals. Each unit cell contains two formula units of BaMX$_3$ and thus has two M atoms, as labeled by A and B, respectively. Under the crystal field of the surrounding MX$_3$ octahedra, the $d$-orbitals of the two M atoms are split into two-fold $e_g$ orbital and three-fold $t_{2g}$ orbitals. The three-fold $t_{2g}$ orbitals is further split into a two-fold $e'_g$ orbitals and a single $a_g$ orbital by the triangle-crystal field of the ionic chain array [see in Fig.\ref{fig:crystal}~(b) for this evolution of $d$-orbitals]. The character of the wave-function mostly composed of $a_g$ manifold is $d_{3z^2-r^2}$ atomic orbital as shown in the inset of Fig.\ref{fig:crystal}~(b). With this quasi-1D crystal structure, the first principles calculations of BaMX$_3$'s electronic structure are performed by using the Vienna \textit{ab initio} simulation package (VASP) \cite{VASP} with generalized gradient approximation \cite{PBE} in the projector augmented-wave method\cite{PAW_Blochl}. The Hubbard $U$ is simulated through Dudarev's method \cite{Dudarev_LDAU} by setting $(U-J)_{\mathrm{V}}= 5.0$ eV and $(U-J)_{\mathrm{Ta}}= 2.0$ eV. Slightly changing of $U-J$ value will not change the conclusion of this work. Tight-binding Hamiltonians are constructed based on the maximally localized Wannier functions (MLWFs)\cite{Vanderbilt_RMP}, and from these Hamiltonians surface band structures are calculated for slabs of BaMX$_3$.

\section{Electronic Structures}
\subsection{Node-Surface in BaVS$_3$}
\begin{figure}[t]
\centering{}
\includegraphics[width=0.5\textwidth]{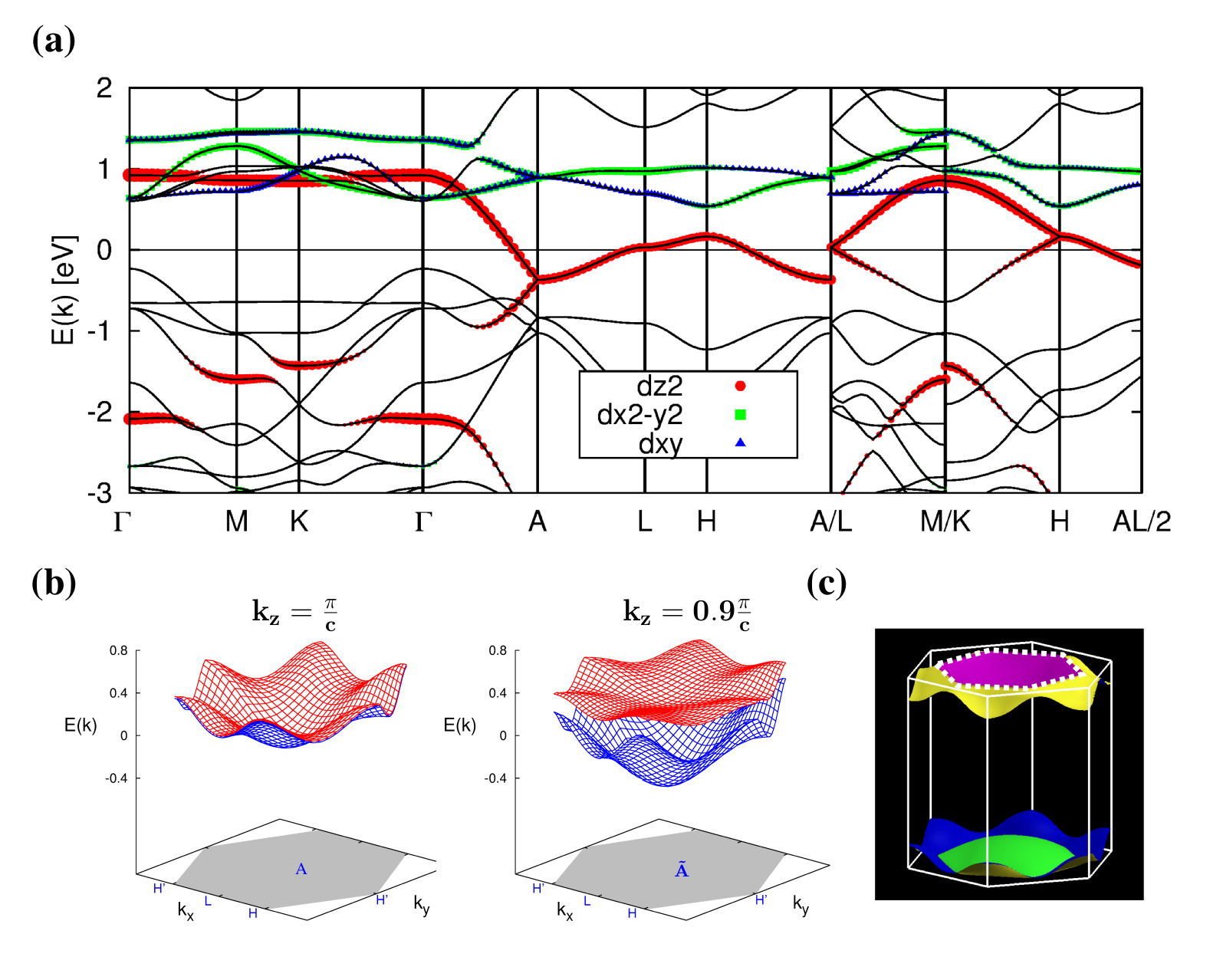}
\caption{ (Color online) Node-surface in BaVS$_3$ under room-temperature paramagnetic phase with negligible spin-orbit coupling. (a) Band dispersions of BaVS$_3$. The orbital nature are represented by circle (red), square (green) and triangle (blue) symbols for the $d_{3z^2-r^2}$,$d_{x^2-y^2}$ and $d_{xy}$ orbitals. (b) 3D plot of the bands in $k_z=\pi/c$ (left) and $k_z=0.9\pi/c$ (right) plane.  (c) Fermi-surface of BaVS$_3$. Dashed lines highlight the contact boundary between two partition of the fermi surface around $k_z=\pi/c$.
\label{fig:nosoc} }
\end{figure}

Firstly, we investigate the electronic structure of BaVS$_3$ in which the effect of SOC can be safely ignored. It is in a paramagnetic phase at room temperature. {It is reported that BaVS$_3$ undergoes a structure phase transition at 240 K and then enters into an insulating state through a metal-insulator phase transition at 70K with the nature of insulating state still being in debate. Here we are only interested in the high temperature paramagnetic state.} The corresponding band structure is shown in Fig.~\ref{fig:nosoc}(a). By analyzing the separated contributions from different $d$-orbitals, we find the electronic states at the fermi-surface are mainly consisted of $a_g$ orbitals. Notably, one can find that the conduction and valence bands exactly stick together at the high-symmetry paths $\mathrm{A-L-H-A}$, while at other directions $\mathrm{\Gamma-M-K-\Gamma-A}$, they are split [see in Fig.\ref{fig:crystal}(c) for the definitions of high-symmetry points]. The dispersions of these bands are small in the $k_z=0$ and $k_z=\pi/c$ planes but large along the $k_z$ direction, indicating the quasi-one-dimensional nature of BaVS$_3$ crystal. By inspecting the band structure along a path off the high symmetry directions, namely $\mathrm{H-{AL}/{2}}$ with $\mathrm{{AL}/{2}}$ being the middle point of path $A-L$, one finds the degeneracy remains and he may guess the degeneracy take place throughout the $k_z=\pi/c$ plane. The conjecture is readily confirmed by the 3D plotting of band structure in Fig.\ref{fig:nosoc} (b). We fix the value of $k_z$ being $\pi/c$ but vary $k_x$ and $k_y$. It is clearly seen that the conduction and valence bands are exactly overlapping at the $k_z=\pi/c$ plane, which is the node-surface with band touching points, or equivalently the Dirac nodal points, in it. Slightly deviating off this plane, $i.e.$, $k_z=0.9\pi/c$, the degeneracy is split. The nodal points are not on the same energy level and 
the Fermi level cuts the node-surface only at a continuous 1D-line on the $k_z=\pi/c$ plane, which connects the hole and electron pockets of Fermi surface indicated with different colors shown in Fig.\ref{fig:nosoc}(c). 

To the best of our knowledge, it is the first time that a node-surface with conduction and valence band-touching nodes is revealed in a realistic material. From band-theory point of view, one would expect that the two $a_g$ orbitals in BaVS$_3$ unit cell may form bonding and anti-bonding bands, and the two valence electrons provided by the V$^{+4}$ ions will fully occupy the bonding band and leave the anti-bonding one empty, making the material a band insulator. This naive band-theory picture, however, fails in the present case due to the nonsymmorphic crystal symmetry contained in the $P6_3/mmc$ group \cite{Parameswaran_NatPhys2015}. The nonsymmorphic symmetry is a crystal symmetry which can not be decomposed into the product of a lattice translation and a point-group operation. {It has long been known\cite{Konig_PRB} that nonsymmorphic symmetries keep energy bands to ``stick together" at the boundary of BZ. The node-surface presented in our work therefore is a faithful demonstration of this physics as checked by the symmetry analysis below.

The participated symmetries to protect the degeneracy at $k_z=\pi/c$ plane are the time-reversal symmetry ${\mathcal{T}}=\mathcal{K}$ with $\mathcal{K}$ being the complex conjugation (in case without spin-oribt cpupling), space inversion $\mathcal{I}$ and the skew axial symmetry $\mathcal{S}_z$= \{$C_{2{z}}|\mathbf{T}_z={\mathbf{c}}/{2}$\}, a nonsymmorphic symmetry combining a half-vector translation along the $\hat{c}$ axis and a two-fold rotation about the $\hat{z}$ axis. From these symmetries, two compound symmetries, $\mathcal{C}=\mathcal{IT}$ and $\mathcal{S}=\mathcal{IS}_z$ are constructed. $\mathcal{C}$ is preserved at any $\mathbf{k}$-point of BZ, while $\mathcal{S}$ is respected at the $k_z=0$ and $k_z=\pi/c$ planes. Applying $\mathcal{S}$ twice on the lattice can bring the system back to its starting point, one has $\mathcal{S}^2=1$ and the corresponding eigen-values of $\mathcal{S}$ are $\pm 1$, which can be used to label the eigen-states of the Hamiltonian. Due to the anti-commutation of $\mathcal{C}$ and $\mathcal{S}$ on $k_z=\pi/c$ plane [see Appendix AI for the derivation], the action of $\mathcal{C}$ switches each eigen-state to its degenerate partner of opposite $\mathcal{S}$-label, ensuring the two-fold degeneracy of energy bands on the entire $k_z=\pi/c$ plane as found in Fig. \ref{fig:nosoc} (b). The details of the symmetry consideration and the derivation of the  actions of the symmetry operators are summarized in Appendix AI.}

\begin{figure}[t]
\centering{}
\includegraphics[width=0.5\textwidth]{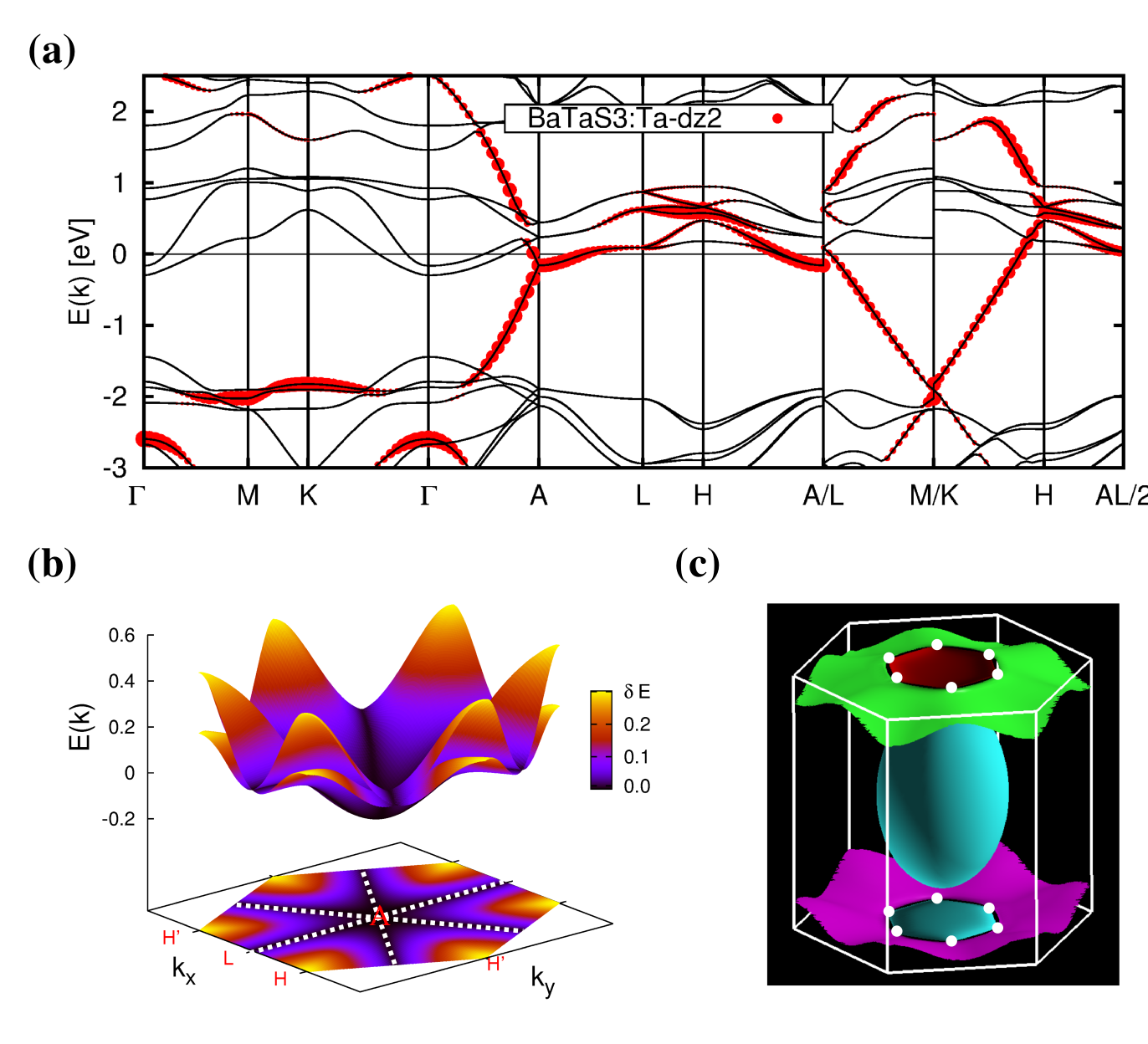}
\caption{ (Color online) Node-lines in BaTaS$_3$ with strong spin-orbit coupling. (a) Band structure of BaTaS$_3$. The orbital characteristic of a$_g$ band is highlighted by circle (red) symbols. (b) 2D plot of the bands in $k_z=\pi/c$ plane. The gradually varied colors indicate the splitting of the bands. The darkest lines along $A-L$ paths are node-lines denoted in the bottom contour plot by dashed lines. (c) Full view of Fermi surface of BaTaS$_3$. The contact points between the two partitions of the fermi surface around $k_z=\pi/c$ are highlighted by solid circles (white).
\label{fig:soc} }
\end{figure}

\subsection{Symmetry Guaranteed Nodal-Lines in BaTaS$_3$}

If SOC is included, the above discussion is not valid and the node-surface of $k_z=\pi/c$ plane will not persist. Luckily for BaVS$_3$, the effective SOC splitting in V's $a_g$ bands is vanishingly small (about $1\sim2$ meV) and thus the conduction and valence bands can still be considered as degenerate at room temperature paramagnetic phase. However, for heavier M = Nb, and Ta, the SOC is large and its effect has to be taken into account. In Fig.\ref{fig:soc}(a), we show the band structure of BaTaS$_3$ with SOC turned on. The inclusion of SOC lifts the degeneracy of bands at a general $\mathbf{k}$-point in the the node-surface $k_z=\pi/c$ plane, as seen from the splitting of energy bands along $\mathrm{L-H-A}$ paths. Surprisingly, one can still find that along a special path $\mathrm{A-L}$, the energy bands remain four-fold degenerate (counting both spin and orbital degrees of freedom), forming a so called node-line structure
with Dirac nodes along it. In Fig.\ref{fig:soc}(b) we also plot a 3D band structure in $k_z=\pi/c$. Gradual colors are used to indicate the energy splitting in the conduction and valence bands. In this plot and its contour projection onto the bottom $k_z=\pi/c$, three node-lines related by $C_{3z}$ rotation symmetry are found along the high-symmetry paths $k_x=0$ and $k_x=\pm{\sqrt{3}}k_y$.

The existence of node-lines also changes the Fermi surface around the $k_z=\pi/c$ plane as shown in Fig.\ref{fig:soc}(c). Due to the non-zero band dispersion of the node-lines, the Fermi level cuts these lines at six discrete points. In the
3D Fermi surface plot, it is readily found that the Fermi lines connecting two different colors in the spinless case [Fig.\ref{fig:nosoc}(c)] now evolve into six separated contacting points on the $A-L$ paths. 
Comparing with the Fermi-surface of BaVS$_3$ in Fig.\ref{fig:nosoc}(c), the additional ellipsoid-shaped Fermi surface surrounding the $\Gamma$ point is from the two $e'_g$ bands as clearly shown in the band structure in Fig.\ref{fig:soc}(a).

Now let us discuss about the symmetry protection of the four-fold degenerate node-lines in presence of SOC. The derivations of the actions of symmetry operators and their associated commutation relations are given in Appendix AII. In the spinful case, time reversal operator is expressed as $\mathcal{T}=is_y\mathcal{K}$ and thus $\mathcal{T}^2=-1$. The skew axis $\mathcal{S}_z$ now also acts on the spin space, $\mathcal{S}_z:(s_x,s_y,s_z)\mapsto(-s_x,-s_y,s_z)$. Applying it twice rotates the spin by $2\pi$, giving a minus sign for a spin-1/2 system. Since acting $\mathcal{S}=\mathcal{IS}_z$ twice on the real space brings the system back to its origin, one finds that the square of $\mathcal{S}$ becomes $\mathcal{S}^2=-1$ at its invariant plane $k_z=\pi/c$. Notice that $\mathcal{I}$ does not act on spin space. Then the eigen-states of Bloch Hamiltonian can be labeled by the eigen-values $\pm i$ of $\mathcal{S}$, $\mathcal{S}|\phi^{\pm}(\mathbf{k})\rangle=\pm i|\phi^{\pm}(\mathbf{k})\rangle$. Unlike the spinless case, applying $\mathcal{C}$ on $|\phi^{\pm}(\mathbf{k})\rangle$ translates it to its Kramer partner with the same $\mathcal{S}$-label. Thus, the two Kramer pairs with opposite $\mathcal{S}$-label are not related and generally the bands should be two-fold degenerate. Therefore, extra symmetries are needed to provide the protection of the node-lines. We find the mirror symmetry $\mathbf{M}_x$ plays this role. $\mathbf{M}_x$ acts both on the real space and spin space. It takes an anti-commutation
relation with $\mathcal{S}$ on the intersection line of its invariant plane $k_x=0$ and the invariant plane of $\mathcal{S}$, $k_z=\pi/c$ . Because of $\{\mathbf{M}_x, \mathcal{S}\}=0$, applying $\mathbf{M}_x$ on $|\phi^{\pm}(\mathbf{k})\rangle$ will translate the state to a degenerate state with opposite $\mathcal{S}$-label. Therefore, with the help of $\mathbf{M}_x$, the two Kramer pairs of opposite $\mathcal{S}$-eigenvalues are now related by $\mathbf{M}_x$ at the high-symmetry path $k_x=0$ on $k_z=\pi/c$, proving the existence of the four-fold degenerate node-lines.

\begin{figure}[t]
\centering{}
\includegraphics[width=0.5\textwidth]{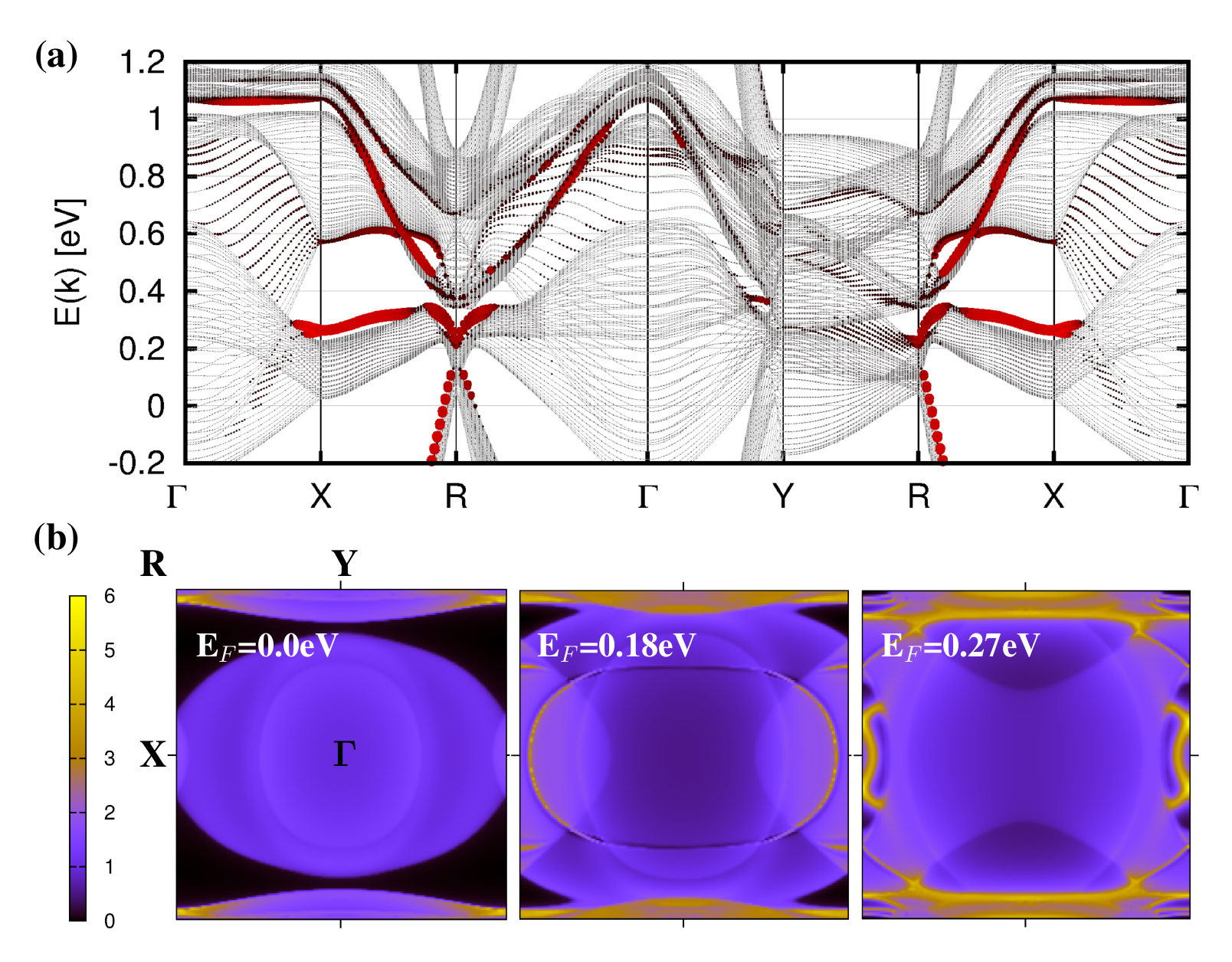}
\caption{ (Color online) Band structure of a 20-unit-cell-thick slab of BaTaS$_3$ with \{-110\} faced.(a) Surface bands highlighted by the thickness and color of the lines scaled by the weight of surface contribution to the eigenstates. (b) Surface local density of states at different Fermi energies, E$_F$ = 0.0 eV (left), 0.18 eV (middle) and 0.27 eV (right).
\label{fig:edge} }
\end{figure}

The node-lines bring about surface states on the surface of BaTaS$_3$, which is another demonstration of the nontrivial topology of the node-line. In Fig.\ref{fig:edge}(a) we plot the surface electronic structure of a 20-unit-cell-thick BaTaS$_3$ slab with \{-110\} facet. From the figure, one can see segments of thick (red) lines inside local ``band gap" which are mainly contributed by the surface layers. When the surface bands immerse into bulk bands, the hybridization between the surface and the bulk bands will smear out the surface contribution. By using recursive Green's function method, we calculated the surface Green's function at the \{-110\} surface and the obtained density of states are plotted in Fig.\ref{fig:edge}(b) for Fermi energy $E_F$ = 0.0, 0.18 and 0.27 eV. Bright lines high-lighting the surface states' contribution can be easily seen in these figures. On the other hand, drumhead-like flat surface bands observed in Graphene-network\cite{WengHM_PRB2015} and Cu$_3$PdN \cite{YuRui_PRL2015}, however, is not found here due to the large dispersions of the node-lines.

The observation of surface states at the slab's surface is unexpected because the symmetries that protect the node-lines are broken at the surfaces. This is understood from the nontrivial Berry phase $\pi$ associated with each node-line\cite{BianG_arxiv2015}. In order to verify this point, we construct an effective four-band Hamiltonian for BaTaS$_3$ in which the two $a_g$ orbitals of Ta are considered [see in Appendix B for the construction of the Hamiltonian]. The low-energy Hamiltonian around a point on the node-line $\mathbf{k}_0=(0,k_{y},\pi/c$) is written as,
\begin{eqnarray}
H_{\mathbf{q}}=\epsilon(\mathbf{k})+q_z\tau_y s_x+q_z\tau_x+q_x\tau_ys_z,
\label{eq:Hq_main}
\end{eqnarray} with $\mathbf{q}=(q_x,q_y,q_z)$ being small deviation from $\mathbf{k}_0$. $\tau$ denotes the orbital degree of freedom and the dependency of $H_{\mathbf{q}}$ on $q_y$ is contained in $\epsilon(\mathbf{k})$ based on the symmetry consideration. As $\epsilon(\mathbf{k})$ is not important for the solutions of (\ref{eq:Hq_main}), we omit it in the following discussion. The effective model (\ref{eq:Hq_main}) has a symmetry $\tau_xs_z$ whose eigenvectors build up a unitary matrix that block-diagonalizes $H_{\mathbf{q}}$. By using an isospin-rotation and re-scaling of $q_z$, one can finally rewrite the Hamiltonian in the nonzero sub-block into the form ${H}_{\tilde{\mathbf{q}}}=\left[\begin{array} {cc} 0  &  \tilde{q}_z+i\tilde{q}_x  \\ \tilde{q}_z-i\tilde{q}_x  &  0\\ \end{array}\right]$. When $\mathbf{\tilde{q}}$ transverses along a closed loop circling the node-line, the wavefunction of the occupied state accumulates a nonzero Berry phase $\pi$ (mod $2\pi$), demonstrating the nontrivial topology of the node-line.

\subsection{Parallel Ionic Chains: Material design for Node-surface and Node-line States}

\begin{figure}[t]
\centering{}
\includegraphics[width=0.5\textwidth]{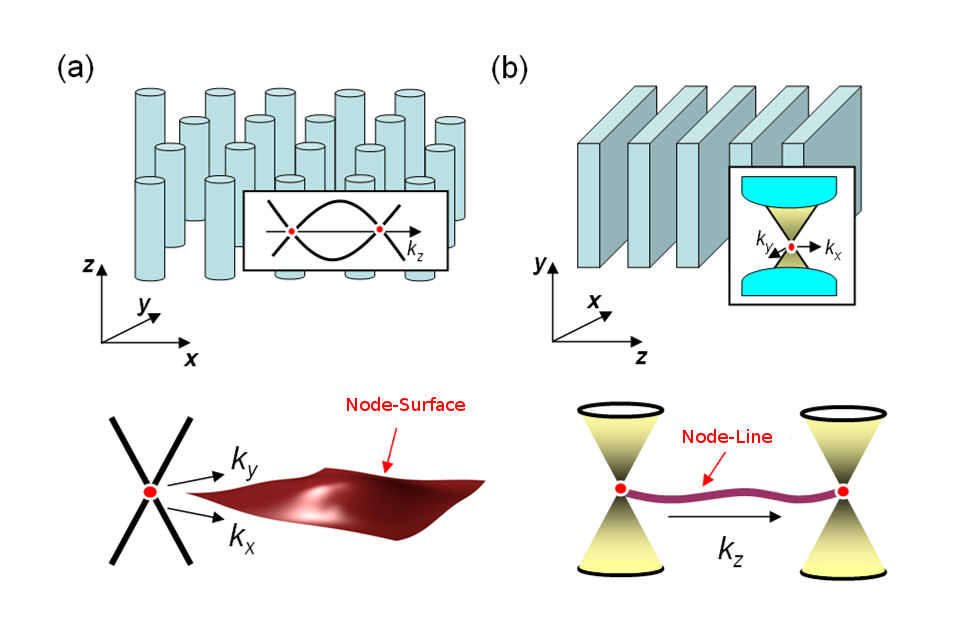}
\caption{ (Color online) Material design for TSMs. (a) (top) parallel arrangement of 1D ionic chains. Inset shows the band structure of a single ionic chains which contains two crossing point at the boundary of BZ, $k_z$=$\pm\pi/c$. (bottom) The two-fold crossing point forms a 2D node-surface degeneracy when the chains are arranged into lattice in the $x$-$y$ plane. (b) (top) Stacking of TI layers to realize a topological semi-metal. Inset shows the 2D Dirac cone band structure of the TI surface state. (bottom) The Dirac point forms a node-line when the layers are stacked up in the $z$ direction.
\label{fig:design} }
\end{figure}

Our work provides a principle for novel materials design: creating TSMs through arrangement of ionic chains in parallel. Previously, Balents \textit{et al.}~\cite{Burkov_PRL2011, burkov} and Phillips \textit{et al.}\cite{Phillips_PRB2014} have considered a layer-stacking approach to construct TSMs. In their scheme, the use of TI layers provides Dirac points from the TI layers' surface states [see in Fig.\ref{fig:design}(b)]. When the TI layers are stacked along the $\hat{z}$ axis, the Dirac points transverses along the $k_z$ direction and forms a node-line in the 3D BZ. The node-line will persist or be split into discrete Weyl points \cite{Burkov_PRL2011} dependent on the details of the inter-layer couplings and the magnetization of the magnetic layers\cite{Phillips_PRB2014}.

Alternatively, here we propose to construct TSMs by parallel-arrangement of ionic chains. The method is schematically shown in Fig. \ref{fig:design}(a). In BaMX$_3$, the individual MX$_3$ chain also respects the skew symmetry $\mathcal{S}_z$, which guarantees an energy band crossing at the BZ's boundary, $k_z=\pm\pi/c$. The parallel-arrangement of the ionic chains then introduces inter-chain coupling. Due to the inter-chain coupling, the crossing point translates along the $k_x$ and $k_y$ directions and builds up a 2D band touching at the 3D BZ's boundary [see the bottom of Fig.\ref{fig:design}(b)]. The inclusion of hopping parameters along $x-y$ plane ususually breaks the degeneracy of the Dirac point at a general $\mathbf{k}$ point where $\mathcal{S}_z$ is broken. Luckily for the present case this nonsymmorphic symmetry is respected at the whole $k_z=\pi/c$ plane and the node-surface persists. When SOC is switched on, the node-surface degeneracy is lifted generally. However, mirror planes containing the $\hat{z}$ axis are presented in the $P6_3/mmc$ group. The node-surface reduces into node-lines at the high-symmetry paths where the mirror planes intersect with node-surface. This is a new route of realizing TSMs with peculiar symmetry-guaranteed node-surface or node-lines.

\section{Discussions and Conclusions}
We notice that the general physics of band-sticking in nonsymmorphic crystals has already been pointed out by Parameswaran \textit{et al.}~\cite{Parameswaran_NatPhys2015} and some first principles calculations have been done on the family of BaMX$_3$\cite{Mattheiss_SSC1995,Whangbo_JSSC2002,JiangXF_PRB2004}. However, the topological features of such 2D degeneracy of conduction and valence bands are revealed for the first time, to the best of our knowledge. The node-lines induced by strong SOC from node-surface without SOC is quite different from those proposed previously\cite{WengHM_PRB2015,Kim_PRL2015,YuRui_PRL2015,XieCava_APLM2015}, in which the node-lines are only stable in the absence of SOC. Some proposals claim the stability of node-lines with the inclusion of SOC, such as those in SrIrO$_3$ \cite{FangC_PRB2015} and PbTaSe$_2$ \cite{BianG_arxiv2015}. These node-lines are off the high-symmetry paths and originate from accidental band touching and no guarantee of their existence. However, the node-lines in the present case are four-fold degenerate and guaranteed by happening at the high-symmetry paths.

We also notice that by decreasing the temperature, BaVS$_3$ experiences firstly a structure phase transition from the room temperature hexagonal phase to the low temperature orthogonal phase, and then a paramagnetic-to-anti-ferromagnetic phase transition under which the material becomes a gapped state. In BaTaS$_3$ and BaTaSe$_3$\cite{Donohue_JSSC1974,Ohtani_MRB2004}, semi-conductor-metal transition are observed at low temperature, however, no clear signature of structure or magnetic transition is detected and the nature of semiconductor-metal transition is still under debate. Whether this phase-transition is topological in nature \cite{Parameswaran_NatPhys2015} or not and what is the relation between the observed phase transitions and the node-surface and node-line discovered in the present work need further investigations.

In summary, we have found a 2D node-surface at the $k_z=\pi/c$ plane in the band structure of BaVS$_3$ with negligible SOC ( about 1$\sim$2 meV). Such novel TSM is protected by the non-symmorphic symmetry $\mathcal{S}_z$. In its family compound BaTaS$_3$ with strong SOC, node-surface evolves into node-lines along three high symmetry paths on the $k_z=\pi/c$ plane. The symmetries that protect the node-lines are the nonsymmorphic skew axial symmetries $\mathcal{S}_z$, mirror symmetry $\mathbf{M}_x$, inversion symmetry $\mathcal{I}$ and time-reversal symmetry $\mathcal{T}$. Two-orbital effective model are constructed to compute the nontrivial Berry phase associated with each node-line. The physics of the node-lines originated from the nonsymmorphic symmetry is analyzed. Surface states are observed in a slab of BaTaS$_3$ even though the large dispersions of node-lines and the breaking of crystal symmetries at the surface. Our work paves a way to design materials with peculiar energy band degeneracy through arrangement of ionic chains in parallel.

\vspace{3mm}
\noindent\textit{Acknowledgements---}
The work is supported by National Natural Foundation of China (NFSC) (Grants No.11574215, No. 11274359 and No. 11422428). Q.F.L acknowledges the support from SRF for ROCS, SEM. H. M. W is also supported by the National 973 program of China (Grants No. 2011CBA00108 and No. 2013CB921700), and the ``Strategic Priority Research Program (B)" of the Chinese Academy of Sciences (Grant No. XDB07020100).

\section{Appendix}
\subsection{Symmetry Analysis}
\subsubsection{The spinless case}
Let us prove that in a crystal preserving the time-reversal-symmetry $\mathcal{T}$, inversion symmetry $\mathcal{I}$ and a two-fold skew-axis along $\hat{z}$, $\mathcal{S}_z:(x,y,z,t)\mapsto (-x,-y,z+\frac{1}{2},t)$, the energy bands at $k_z=\pi/c$ plane will be two-fold degenerated if the spin effect is excluded. By using these three symmetries, one can compose two compound symmetries, $\mathcal{C}=\mathcal{IT}$ and $\mathcal{S}=\mathcal{IS}_z$. In the real space, $\mathcal{C}$ and $\mathcal{S}$ act as,
\begin{eqnarray}
\mathcal{C}:(x,y,z,t)\mapsto(-x,-y,-z,-t)\nonumber\\
\mathcal{S}:(x,y,z,t)\mapsto(x,y,-z-\frac{1}{2},t).\nonumber\\
\label{eq:CS}
\end{eqnarray}
In momentum space, one easily finds that all the $\mathbf{k}$ points are invariant points under the action of $\mathcal{C}$. On the other hand, $\mathcal{S}$ translates $(k_x,k_y,k_z)$ to $(k_x,k_y,-k_z)$, which means $k_z=0$ and $k_z=\pi/c$ are the two invariant planes of $\mathcal{S}$.
Applying $S$ twice, the coordinates in the real space return back and one finds $\mathcal{S}^2=1$, which means the eigen-values of $\mathcal{S}$ are $\pm 1$ and one can use them to label the Bloch states $|\psi(\mathbf{k})\rangle$ of the system's Hamiltonian by $\mathcal{S}|\psi^{\pm}(\mathbf{k})\rangle=\pm|\psi^{\pm}(\mathbf{k})\rangle$.

 We find that under the action of $\mathcal{C}$, each eigen-state $|\psi^{\pm}(\mathbf{k})\rangle$ is switched to a degenerate state of an opposite $\mathcal{S}$-label. Before showing this, we first prove that $\mathcal{C}$ and $\mathcal{S}$ take an anti-commutation relation at $k_z=\pi/c$. By using Eq.(\ref{eq:CS}), one obtain,
\begin{eqnarray}
\mathcal{SC}:(x,y,z,t)\mapsto(-x,-y,z-\frac{1}{2},-t)\nonumber\\
\mathcal{CS}:(x,y,z,t)\mapsto(-x,-y,z+\frac{1}{2},-t),
\label{eq:CS_SC}
\end{eqnarray} from which one finds in $\mathbf{k}$-space,
\begin{eqnarray}
\mathcal{CS}=\mathbf{T}_{\mathbf{c}}\mathcal{SC}=e^{i{k_z}{c}}\mathcal{SC},
\label{eq:CS-SC}
\end{eqnarray} with $\mathbf{T}_{\mathbf{c}}$ being the translation along the $\hat{z}$ axis by one primary vector $\mathbf{c}$. Therefore the anti-commutator of $\mathcal{C}$ and $\mathcal{S}$ becomes zero, $\{\mathcal{C},\mathcal{S}\}=0$ at $k_z=\pi/c$. With this anti-commutation relation, one can prove that $\mathcal{C}$ relates the eigenstates to their degenerate partners of opposite $\mathcal{S}$-label as follows,
\begin{eqnarray}
\mathcal{S}\mathcal{C}|\psi^{\pm}(\mathbf{k})\rangle=e^{-i\frac{\pi}{c}c}\mathcal{C}\mathcal{S}|\psi^{\pm}(\mathbf{k})\rangle=\mp\mathcal{C}|\psi^{\pm}(\mathbf{k})\rangle.
    \label{eq:label}
\end{eqnarray}
Then on the $k_z=\pi/c$ plane, one proves that each band is two-fold degenerated. We should stress here that for the case of no spin, actually the inversion $\mathcal{I}$ is not necessary for the degeneracy at $k_z=\pi/c$. We can use an alternative compound symmetry $\mathbf{\Theta}=\mathcal{TS}_z$ to protect the degeneracy. It is easy to check that it takes $\mathbf{\Theta}^2=-1$ at $k_z=\pi/c$ plane, which ensures a Krammer degeneracy with the two related states labeled by eigen-values $\pm i$ of $\mathbf{\Theta}$. However, in the following spinful case, $\mathbf{\Theta}$ is not enough to protect the node-lines. For the purpose of consistence with the spinful case below, we use the same compound symmetries $\mathcal{IS}_z$ and $\mathcal{IT}$ in the spinless case.

\subsubsection{the effect of spin-orbit coupling}

Including the spin degree of freedom and the effect of SOC generally splits the four-fold degeneracy (counting both spin and orbit degree of freedom) at $k_z=\pi/c$. In this spinful case, $\mathcal{S}_z$ also acts on the
spin space, $\mathcal{S}_z:(s_x,s_y,s_z)\mapsto(-s_x,-s_y,s_z)$. The square of $\mathcal{S}_z$ then rotates the spins by $2\pi$, contributing a minus sign for the spin$-\frac{1}{2}$ system. Therefore the square of $\mathcal{S}=\mathcal{IS}_z$ becomes $\mathcal{S}^2=-1$ and $\mathcal{S}$ now has eigen-values $\pm i$, which again can be used to label the eigen-states of Bloch Hamiltonian, $\mathcal{S}|\phi^{\pm}(\mathbf{k})\rangle=\pm i|\phi^{\pm}(\mathbf{k})\rangle$. Unlike the spinless case, the action of $\mathcal{C}=\mathcal{IT}$ on the eigen-states $|\phi^{\pm}(\mathbf{k})\rangle$ only switches it to a partner state of the same $\mathcal{S}$-label, even though the anti-commutation relation of $\mathcal{C}$ and $\mathcal{S}$ retains. This point is easily checked as,
\begin{eqnarray}
\mathcal{S}\mathcal{C}|\phi^{\pm}(\mathbf{k})\rangle&=&e^{-ik_zc}\mathcal{C}\mathcal{S}|\phi^{\pm}(\mathbf{k})\rangle\nonumber \\
    &=& -\mathcal{C}\left(\pm i|\phi^{\pm}(\mathbf{k})\rangle \right)\nonumber\\
    &=& \pm i \mathcal{C}|\phi^{\pm}(\mathbf{k})\rangle.
    \label{eq:label_soc}
\end{eqnarray}
Eq.(\ref{eq:label}) is used in the first line of Eqs.(\ref{eq:label_soc}) and the last equality in Eq.(\ref{eq:label_soc}) is obtained based on that time reversal operator $\mathcal{T}=i s_y\mathcal{K}$ contained in $\mathcal{C}$ conjugates $i$ to $-i$.

As $\mathcal{C}$ only related the eigen-state to its partner with the same $\mathcal{S}$-label, we only obtain two Krammers pairs that can not be related by symmetry $\mathcal{S}$ and $\mathcal{C}$ only.
To explain the four-degenerated node-lines of BaTaS$_3$, we need other symmetries. We find it is the mirror symmetry $\mathbf{M}_x:(x,y,z)\mapsto(-x,y,z)$ that provides the needed protection for these node-lines. The implication is that the high symmetry path $k_x=0$ at $k_z=\pi/c$ is just the intersecting line of the invariant plane $k_x=0$ of $M_x$ and the invariant plane $k_z=\pi/c$ of $\mathcal{S}$. On this invariant line, action of $\mathbf{M}_x$ will transform the eigen-state $|\phi^{\pm}(\mathbf{k})\rangle$ to a partner of an opposite $\mathcal{S}$-label, which then related the two Krammer pairs and ensures a four-fold degenerate node-line. Before clarifying this point, we first prove that $\mathbf{M}_x$ anti-commutes with $\mathcal{S}$. It is easily checked that in the real space, action of $\mathcal{S}\mathbf{M}_x$ and $\mathbf{M}_x\mathcal{S}$ on the coordinate lead to the same result. The anti-commutation relation of $\mathbf{M}_x$ and $\mathcal{S}$ comes from their action on the spin space. In spin space, $\mathcal{S}$ reflects $s_x$ and $s_y$ but keeps $s_z$ invariant, from which one can express its action by $\exp\left(i\pi/2\hat{s}_z\right)=i\hat{s}_z$. Similarly, the action of $\mathbf{M}_z$ on the spin is expressed by $\exp\left(i\pi/2\hat{s}_x\right)=i\hat{s}_x$. It is the action of $\mathbf{M}_x$ and $\mathcal{S}$ on the spin space that ensure the anti-commutation relation, $\{\mathcal{S},\mathbf{M}_x\}=0$.  By applying $\mathbf{M}_x$ on the eigenstate $|\phi^{\pm}(\mathbf{k})\rangle$, one then finds,
\begin{eqnarray}
\mathcal{S}\mathbf{M}_x|\phi^{\pm}(\mathbf{k})\rangle&=&-\mathbf{M}_x\mathcal{S}|\phi^{\pm}(\mathbf{k})\rangle \nonumber\\
                                                     &=&-\mathbf{M}_x\left(\pm i|\phi^{\pm}(\mathbf{k})\rangle\right)\nonumber\\
                                                     &=&\mp i\mathbf{M}_x|\phi^{\pm}(\mathbf{k})\rangle.
    \label{eq:label_m}
\end{eqnarray}
Eqs.(\ref{eq:label_soc}) and (\ref{eq:label_m}) finally finish the proving that at $k_x=0$ on the $k_z=\pi/c$ plane there must be a four-fold-degenerated node-line guaranteed by symmetries $\mathcal{S}=\mathcal{IS}_z$, $\mathcal{C}=\mathcal{IT}$ and $\mathbf{M}_x$. The node-lines at $k_x=\pm\sqrt{3}k_y$ on the $k_z=\pi/c$ are then resulted from the $C_{3z}$ rotation.

\subsection{Effective Hamiltonian}
There are two active d${3z^2-r^2}$ orbits hosted on Ta atom in the unit cell and they compose the electronicstates around the fermi surface. When SOC is turned off, we can use these two orbits, denoted by A and B, to construct a $2\times2$ Hamiltonian at each $\mathbf{k}$-point in the momentum space,
\begin{eqnarray}
\hat{H}_0(\mathbf{k})=\left[
\begin{array}{cc}
H_{AA}(\mathbf{k}) & H_{AB}(\mathbf{k}) \\
H_{AB}^{*}(\mathbf{k}) &H_{BB}(\mathbf{k}) \\
\end{array}
 \right].\label{eq:H_0}
\end{eqnarray}
Here $|\eta,\mathbf{k}\rangle=\frac{1}{\sqrt{N}}\sum_n e^{i \mathbf{k} \cdot\mathbf{r}_{n\eta}}\psi(\mathbf{r}-\mathbf{r}_{n\eta})$ are the two basis functions with $\eta=A,B$ denoting the orbits. $\psi(\mathbf{r}-\mathbf{r}_{n\eta})$ is the local $d3z^2-r^2$ wave function centered on atom site $\mathbf{r}_{n\eta}$ in the n$^{th}$ unit cell. For the spinful Hamiltonian, We need to take the spin degree of freedom into consideration and thus define four basis functions, $|\eta,\mathbf{k},\sigma\rangle$ with $\sigma=\uparrow $ or $\downarrow$ denoting the spin direction. The Hamiltonian is then enlarged to a $4\times4$ size,
\begin{eqnarray}
H(\mathbf{k})=\left[
\begin{array} {cccc}
H_{AA}^{\uparrow\uparrow}(\mathbf{k}) &H_{AB}^{\uparrow\uparrow}(\mathbf{k}) &H_{AA}^{\uparrow\downarrow}(\mathbf{k}) &H_{AB}^{\uparrow\downarrow}(\mathbf{k}) \\
 &H_{BB}^{\uparrow\uparrow}(\mathbf{k}) &H_{BA}^{\uparrow\downarrow}(\mathbf{k}) &H_{BB}^{\uparrow\downarrow}(\mathbf{k}) \\
 &  & H_{AA}^{\downarrow\downarrow}(\mathbf{k}) &H_{AB}^{\downarrow\downarrow}(\mathbf{k}) \\
\dag  &  &  & H_{BB}^{\downarrow\downarrow}(\mathbf{k})
\label{eq:H}
\end{array}
\right].\nonumber\\
\end{eqnarray}

The time-reversal operator $\mathcal{T}$ now becomes
$\mathcal{T}=i\hat{s}_y\mathcal{K}$ with $\hat{s}_y=\left[\begin{array}{cc} 0& i\\-i& 0 \end{array}\right]$ being the Pauli matix acting on the spin space. It reads,
\begin{eqnarray}
\mathcal{T}|\eta,\mathbf{k},\sigma \rangle  & = & \mathrm{sng}(\sigma) |\eta,-\mathbf{k},\bar{\sigma} \rangle  \nonumber\\
\mathcal{T}H(\mathbf{k}) & = & H(-\mathbf{k})\mathcal{T}.
\label{eq:T_psi2}
\end{eqnarray} with the sign function sng$(\sigma)$ being = $+$1 ($-$1) for spin $\uparrow$ ($\downarrow$) and $H(\mathbf{k})=e^{-i\mathbf{k}\cdot\mathbf{r}}H(\mathbf{r})e^{i\mathbf{k}\cdot\mathbf{r}}$.
For the entries of Hamiltonian (\ref{eq:H}), one finds,
\begin{eqnarray}
&H&_{\eta_1\eta_2}^{\sigma_1\sigma_2} (\mathbf{k})  =   \langle \eta_1,\mathbf{k},\sigma_1| H(\mathbf{k}) | \eta_2,\mathbf{k}, \sigma_2\rangle  \nonumber \\
   & =&  -\mathrm{sgn}(\sigma_2) \langle\eta_1,\mathbf{k},{\sigma}_1| H(\mathbf{k}) \mathcal{T} | \eta_2,-\mathbf{k}, \bar{\sigma}_2\rangle  \nonumber \\
    &= & -\mathrm{sgn}(\sigma_2) \langle\eta_1,\mathbf{k},{\sigma}_1|\mathcal{T} \left[H(\mathbf{-k}) | \eta_2,-\mathbf{k}, \bar{\sigma}_2\rangle \right] \nonumber \\
    &=&  \mathrm{sgn}(\sigma_2)\mathrm{sgn}(\sigma_1) \langle\mathcal{T} (\eta_1,-\mathbf{k},\bar{\sigma}_1)|\mathcal{T} \left[H(-\mathbf{k}) | \eta_2,-\mathbf{k}, \bar{\sigma}_2\rangle \right] \nonumber \\
    &=&  \mathrm{sgn}(\sigma_1)\mathrm{sgn}(\sigma_2) \left[\langle\eta_2,-\mathbf{k},\bar{\sigma}_2| H(-\mathbf{k})\right] | \eta_1,-\mathbf{k}, \bar{\sigma}_1\rangle \nonumber \\
    &=&  \mathrm{sgn}(\sigma_1)\mathrm{sgn}(\sigma_2) H_{\eta_2\eta_1}^{\bar{\sigma}_2\bar{\sigma}_1} (-\mathbf{k}).
\label{eq:T_spin}
\end{eqnarray}
To obtain the second and forth lines of the above equation, we use the transformation (\ref{eq:T_psi2}) of wave function for the right and left ket, respectively. The transformation of Hamiltoina under $\mathcal{T}$ is used to obtain the third line. To get the fifth line, we use the property of time-reversal symmetry $\langle \mathcal{T}\phi|\mathcal{T}|\psi\rangle=\langle\psi|\phi\rangle$.

The $P6_3/mmc$ group also contains an inversion symmetry $\mathcal{I}:(x,y,z)\mapsto(-x,-y,-z)$. ${\mathcal{I}}$ does not flip the spin and orbit indices, and the wave functions and Hamiltonian are transformed as,
\begin{eqnarray}
{\mathcal{I}}|\eta,\mathbf{k},\sigma \rangle  & = &  |\eta,-\mathbf{k}, {\sigma} \rangle  \nonumber\\
{\mathcal{I}}H(\mathbf{k}) & = & H(-\mathbf{k}){\mathcal{I}}.
\label{eq:I_psi}
\end{eqnarray}
Applying Eq. (\ref{eq:I_psi}) one finds the consequence of applying $\mathcal{I}$ onto the elements of Hamiltonian (\ref{eq:H}),
\begin{eqnarray}
H_{\eta_1\eta_2}^{\sigma_1\sigma_2}(k_x,k_y,k_z)=H_{ {\eta}_1 {\eta}_2}^{\sigma_1\sigma_2}(-k_x,-k_y,-k_z).
\label{eq:I_spin}
\end{eqnarray}

Applying Eqs.(\ref{eq:T_spin}) and (\ref{eq:I_spin}) successively one obtains
\begin{equation}
H_{\eta_1\eta_2}^{\sigma_1\sigma_2}(k_x,k_y,k_z)=\mathrm{sng}(\sigma_1)\mathrm{sng}({\sigma}_2)H_{\eta_2\eta_1}^{\bar{\sigma}_2\bar{\sigma}_1}(k_x,k_y,k_z) ,
\label{eq:IT}
\end{equation}
 which reduces the Hamiltonian (\ref{eq:H}) to the below form,
\begin{eqnarray}
H({\mathbf{k}})=\left[
\begin{array} {cccc}
a({\mathbf{k}}) &  f({\mathbf{k}}) & 0 &g({\mathbf{k}}) \\
   &b({\mathbf{k}}) & -g({\mathbf{k}})  & 0  \\
   &  & a({\mathbf{k}}) &  f^{*}({\mathbf{k}})\\
 \dag  & &  & b({\mathbf{k}})
\end{array}
\right],
\label{eq:H_IT}
\end{eqnarray} where $a({\mathbf{k}})$ and $b({\mathbf{k}})$ are real functions, and $f({\mathbf{k}})$ and $g({\mathbf{k}})$ are complex functions.

Hamiltonian (\ref{eq:H_IT}) is mathematically  equivalent to a linear combination of five Dirac matrices $\Gamma_{a}$ together with the unit matrix $\Gamma_0$. The eigenvalues of Hamiltonian (\ref{eq:H_IT}) become four-fold
degenerate only if the five coefficients of the Dirac matrices become zero, that is to say $a(\mathbf{k})-b(\mathbf{k}) = f(\mathbf{k}) = g(\mathbf{k}) =0$, which, however, does not hold at a general $\mathbf{k}$ point. Below we prove that this four-fold-degenerating condition readily fulfills at three high-symmetries paths, $k_x=0$ and $k_x=\pm{\sqrt{3}}k_y$ on the $k_z=\pi/c$ plane with the help of $\mathcal{S}_z$ and a mirror symmetry $\textbf{M}_x:(x,y,z)\mapsto(-x,y,z)$.

In real space, the skew axial symmetries $\mathcal{S}_z$ flips the orbit indices. While in spin space, it keeps the $s_z$ component invariant but reverse $s_x$ and $s_y$, indicating that its action can be represented by $i\hat{s}_z$. Therefore the wave function and Hamiltonian transform as,
\begin{eqnarray}
\mathcal{S}_z|\eta,\mathbf{k},\sigma \rangle  & = & ie^{-i\frac{k_zc}{2}}\mathrm{sng}(\sigma) |\bar{\eta},-k_x,-k_y,k_z,{\sigma} \rangle  \nonumber\\
\mathcal{S}_zH(\mathbf{k}) & = & H(-k_x,-k_y,k_z)\mathcal{S}_z.
\label{eq:Sz_psi1}
\end{eqnarray}

With Eq.(\ref{eq:Sz_psi1}), one finds the below relation for Bloch Hamiltonian (\ref{eq:H_IT}),
\begin{eqnarray}
H_{\eta_1\eta_2}^{\sigma_1\sigma_2}(\mathbf{k}) & = & \langle \eta_1,\mathbf{k},\sigma_1| {H}(\mathbf{k}) | \eta_2,\mathbf{k},\sigma_2\rangle \nonumber\\
   & = &   \langle \eta_1,\mathbf{k},\sigma_1| \hat{\mathcal{S}_z}^{-1} \hat{\mathcal{S}_z}{H}(\mathbf{k})\hat{\mathcal{S}_z}^{-1} \mathcal{S}_z| \eta_2,\mathbf{k},\sigma_2\rangle \nonumber\\
   & = &   \langle \eta_1,\mathbf{k},\sigma_1| \hat{\mathcal{S}_z}^{-1} {H}(\tilde{\mathbf{k}}) \hat{\mathcal{S}_z} | \eta_2,\mathbf{k},\sigma_2\rangle \nonumber\\
   & = &   \langle \bar{\eta}_1,\tilde{\mathbf{k}},\sigma_1| {H}(\tilde{\mathbf{k}}) | \bar{\eta}_2,\tilde{\mathbf{k}},\sigma_2\rangle \nonumber\\
   & = & \mathrm{sng}(\sigma_1)  \mathrm{sng}(\sigma_2)H_{\bar{\eta}_1\bar{\eta}_2}^{\sigma_1\sigma_2} (-k_x,-k_y,k_z),
\label{eq:S_aa}
\end{eqnarray} with $\tilde{\mathbf{k}}=(-k_x,-k_y,k_z)$.
Then a constraint for the spinful Hamiltonian (\ref{eq:H_IT}) is obtained,
\begin{equation}
H_{\eta_1\eta_2}^{\sigma_1\sigma_2}(k_x,k_y,k_z)=\mathrm{sng}(\sigma_1)\mathrm{sng}(\sigma_2)H_{\bar{\eta}_1\bar{\eta}_2}^{\sigma_1\sigma_2}(-k_x,-k_y,k_z).
\label{eq:Skew_XZ}
\end{equation}
 Combining Eqs.(\ref{eq:I_spin}) and (\ref{eq:Skew_XZ}), one arrives at the below important relation of Hamiltonian posed by symmetry $\mathcal{S}=\mathcal{IS}_z$,
\begin{equation}
H_{\eta_1\eta_2}^{\sigma_1\sigma_2}(k_x,k_y,k_z)=\mathrm{sgn}(\sigma_1)\mathrm{sgn}(\sigma_2)H_{\bar{\eta}_1\bar{\eta}_2}^{{\sigma}_1{\sigma}_2}(k_x,k_y,-k_z).
\label{eq:Skew_final}
\end{equation}

Applying the transformation $H_{\eta_1\eta_2}^{\sigma_1\sigma_2}(\mathbf{k}+\mathbf{G})$ = $e^{i\mathbf{G}\cdot\mathbf{r}_{\eta_1\eta_2}}H_{\eta_1\eta_2}^{\sigma_1\sigma_2}(\mathbf{k})$ to Eq.(\ref{eq:Skew_final}) for $k_z=\pi/c$, one finds 
\begin{eqnarray}
&&H_{AB}^{\sigma_1\sigma_2}(k_x,k_y,\pi/c)  \nonumber \\
&=&\mathrm{sgn}(\sigma_1)\mathrm{sgn}(\sigma_2)H_{BA}^{{\sigma}_1{\sigma}_2}(k_x,k_y,-\pi/c) \nonumber \\
&=& \mathrm{sgn}(\sigma_1)\mathrm{sgn}(\sigma_2) e^{i\frac{2\pi}{c}\frac{c}{2}}H_{BA}^{{\sigma}_1{\sigma}_2}(k_x,k_y,\pi/c),
\label{eq:cons1}
\end{eqnarray}
so that 
\begin{eqnarray}
H_{AB}^{\uparrow\downarrow}(k_x,k_y,\pi/c)&=&H_{BA}^{\uparrow\downarrow}(k_x,k_y,\pi/c) \nonumber\\
H_{AB}^{\uparrow\uparrow}(k_x,k_y,\pi/c)&=&-H_{BA}^{\uparrow\uparrow}(k_x,k_y,\pi/c), 
\label{eq:fg}
\end{eqnarray}
proving in Hamiltonian (\ref{eq:H_IT}) that 
\begin{eqnarray}
g(k_x,k_y,\pi/c)=0,~~\mathbf{Re}\left[f(k_x,k_y,\pi/c)\right]=0.
\label{eq:g_ref}
\end{eqnarray}
Applying Eq. (\ref{eq:Skew_final}) to $H_{AA}^{\uparrow\uparrow}(\mathbf{k})$, one similarly obtains 
\begin{eqnarray}
H_{AA}^{\uparrow\uparrow}(k_x,k_y,\pi/c)=H_{BB}^{\uparrow\uparrow}(k_x,k_y,\pi/c),
\label{eq:cons2}
\end{eqnarray}
 which proves 
\begin{eqnarray}
a(k_x,k_y,\pi/c) = b(k_x,k_y,\pi/c).
\label{eq:ab}
\end{eqnarray}
Therefore, in the presence of $\mathcal{I}, \mathcal{T}$ and $\mathcal{S}_z$, most elements of Hamiltonian (\ref{eq:H_IT}) vanishes at $k_z=\pi/c$ plane except imaginary party of $H_{AB}^{\uparrow\downarrow}(k_x,k_y,\pi/c)$ and a diagonal element $\epsilon(\mathbf{k})=[a(k_x,k_y,\pi/c)+b(k_x,k_y,\pi/c)]/{2}$.

It is possible to prove that $\mathbf{Im}\left[f(0,k_y,\pi/c)\right]=0$ at the high-symmetry path $k_x=0$ on $k_z=\pi/c$ plane. To do so, we need extra symmetries, which in the present case is a mirror symmetry $\mathbf{M}_x$ with the mirror plane lying at the $y$-$z$ plane.
In real space the mirror symmetry $\mathbf{M}_x$ keeps the orbit indices invariant and inverses the $x$-component of coordinates. While in spin space, it keeps the spin component $s_x$ invariant but reverse $s_y$ and $s_z$, indicating its action on the spin space represented by $i\hat{s}_x$. Therefore the wave function and Hamiltonian transform as,
\begin{eqnarray}
\mathbf{M}_x|\eta,\mathbf{k},\sigma \rangle  & = & i|{\eta},-k_x,k_y,k_z,\bar{\sigma} \rangle  \nonumber\\
\mathbf{M}_xH(\mathbf{k}) & = & H(-k_x,k_y,k_z)\mathbf{M}_x.
\label{eq:Sz_psi2}
\end{eqnarray}
which indicates,
\begin{eqnarray}
H_{\eta_1\eta_2}^{\sigma_1\sigma_2}(k_x,k_y,k_z)= H_{{\eta}_1 {\eta}_2}^{\bar{\sigma}_1\bar{\sigma}_2}(-k_x,k_y,k_z).
\label{eq:mx_spin}
\end{eqnarray}
With Eq.(\ref{eq:mx_spin}), one gets $H_{AB}^{\uparrow\uparrow}(0,k_y,k_z)= H_{AB}^{\downarrow\downarrow}(0,k_y,k_z)$ and $H_{AB}^{\uparrow\downarrow}(0,k_y,k_z)= H_{AB}^{\downarrow\uparrow}(0,k_y,k_z)$, which proves
\begin{eqnarray}
\mathbf{Im}\left[f(0,k_y,k_z)\right]=0, ~\mathbf{Re}\left[g(0,k_y,k_z)\right]=0.
\label{eq:f_reg}
\end{eqnarray}
For entry $H_{AA}^{\uparrow\uparrow}$, $\mathbf{M}_x$ also sets $H_{AA}^{\uparrow\uparrow}(0,k_y,k_z)=H_{AA}{\downarrow\downarrow}(0,k_y,k_z)$ which results into the constraint,

 By summarizing the above discussion, we prove that there exists a node-line at the $k_x=0$ path on $k_z=\pi/c$ plane for BaMX$_3$ when SOC is turned on. For other two
node-lines at $k_x=\pm\sqrt{3}k_y$ (see in Fig.~\ref{fig:soc}(b)), they can be produced by using the $\hat{C}_3$ symmetry of the system, which rotates the $k_x=0$ node-line to the $k_x=\pm\sqrt{3}k_y$ directions.

We then expand the Hamiltonian around a point on the node-line $\mathbf{k}=(0,k_y,\pi/c)+(q_x,q_y,q_z)$ with $(q_x,q_y,q_z)$ being small deviations from the expanding point $\mathbf{k}_0=(0,k_y,\pi/c)$. 
 From the above Eqs. (\ref{eq:g_ref}), (\ref{eq:ab}) and (\ref{eq:f_reg}) one finds,
\begin{eqnarray}
\frac{\partial g}{\partial q_x}\Big|_{\mathbf{k}_0}&=&0,~\frac{\partial g}{\partial q_y}\Big|_{\mathbf{k}_0}=0,~\mathbf{Re}\frac{\partial g}{\partial q_z}\Big|_{\mathbf{k}_0}=0,\nonumber\\
\mathbf{Re}\frac{\partial f}{\partial q_x}\Big|_{\mathbf{k}_0}&=&0,~\frac{\partial f}{\partial q_y}\Big|_{\mathbf{k}_0}=0,~\mathbf{Im}\frac{\partial f}{\partial q_z}\Big|_{\mathbf{k}_0}=0 \nonumber\\
\frac{\partial a}{\partial q_i}\Big|_{\mathbf{k}_0}&=&\frac{\partial b}{\partial q_i}\Big|_{\mathbf{k}_0}, (i=x,y,z)
\label{eq:H_expand}
\end{eqnarray}
from which the low energy Hamiltonian is obtained,
\begin{eqnarray}
H_{\mathbf{q}}=\epsilon(\mathbf{k})+\alpha q_z\tau_y s_x+ q_z\tau_x+q_x\tau_ys_z
\label{eq:Hq}
\end{eqnarray}
As $\epsilon(\mathbf{k})$ is not important for the solution of Hamiltonian (\ref{eq:H_expand}), we omit it in the following discussions. The Hamiltonian (\ref{eq:Hq}) takes a symmetry $\tau_xs_y$ and by using the unitary matrix $U$ which diagonalizes $\tau_xs_y$,
 \begin{eqnarray}
U=\frac{1}{\sqrt{2}}\left[
\begin{array} {cccc}
 1  &  0  &  1  &  0 \\
 0  &  1  &  0 &  1  \\
 0  &  -i  &  0  &  i  \\
 -i  &  0  &  i  & 0  \\
\end{array}
\right],
\label{eq:U}
\end{eqnarray}
one can transform the Hamiltonian $H_{\mathbf{q}}$ into a block-diagonalized matrix $\tilde{H}_{\mathbf{q}}$,
 \begin{eqnarray}
\tilde{H}_{\mathbf{q}}=\epsilon(\mathbf{k})+\frac{1}{\sqrt{2}}\left[
\begin{array} {cccc}
 \alpha q_z  &  q_z+iq_x  &  0  &  0 \\
 q_z-iq_x  &  -\alpha q_z  &  0 &  0  \\
 0  &  0  &  -\alpha q_z  &  q_z+iq_x  \\
 0  &  0  &  q_z-iq_x  & \alpha q_z  \\
\end{array}
\right].\nonumber \\
\label{eq:H_q2}
\end{eqnarray}
For each nonzero sub-block of Hamiltonian (\ref{eq:H_q2}), the diagonal terms are further moved to the off-diagonal entries through an isospin-rotation about the $\tilde{\tau}_y$ axis $\exp(i{\pi}/{8}\tilde{\tau}_y)$. After a rescale of $q_z$ by $1/(\alpha^2+1)$ , one obtains a simple form of $2\times2$ Hamiltonian ${H}_{\tilde{\mathbf{q}}}=\left[\begin{array} {cc} 0  &  \tilde{q}_z+i\tilde{q}_x  \\ \tilde{q}_z-i\tilde{q}_x  &  0\\ \end{array}\right]$ for each sub-block. The occupied state has an eigen-value of $-\sqrt{\tilde{q}_x^2+\tilde{q}_z^2}$ with the corresponding eigen-vector being ${\sqrt{2}}/{2}\left[\begin{array} {c} 1 \\  -\exp(-i\theta) \end{array}\right]$. When $\mathbf{\tilde{q}}$ transverses along a closed loop circling the node-line, the accumulated Berry phase is computed to $\pi$ as follows,
\begin{eqnarray}
\mathbf{\Phi}_B&=&-i\lim_{N\rightarrow\infty}\sum_{j=0}^{N-1}\log\langle j|j+1\rangle\nonumber \\
&=& -i\lim_{N\rightarrow\infty}\sum_{j=0}^{N-1}\log\left(\frac{1+\exp[i(\theta_{j+1}-\theta_j)]}{2}\right)\nonumber \\
&=&-i\int_0^{2\pi}\frac{id\theta}{2}=\pi,
\label{eq:Berry}
\end{eqnarray} where the loop is discretized into N successive points labeled by $j$ $(j=0,1,2,\ldots,N-1)$ with $N=0$. The Berry phase $\mathbf{\Phi}_B$ here is only meaningful for modulo $2\pi$ as $|N\rangle=e^{i2m\pi}|0\rangle$ is also well defined for condition $N=0$, which will increase $\mathbf{\Phi}_B$ by $2m\pi$.

\subsection{Tight-Binding Model}

\begin{figure}[!thb]
\centering{}
\includegraphics[width=0.5\textwidth]{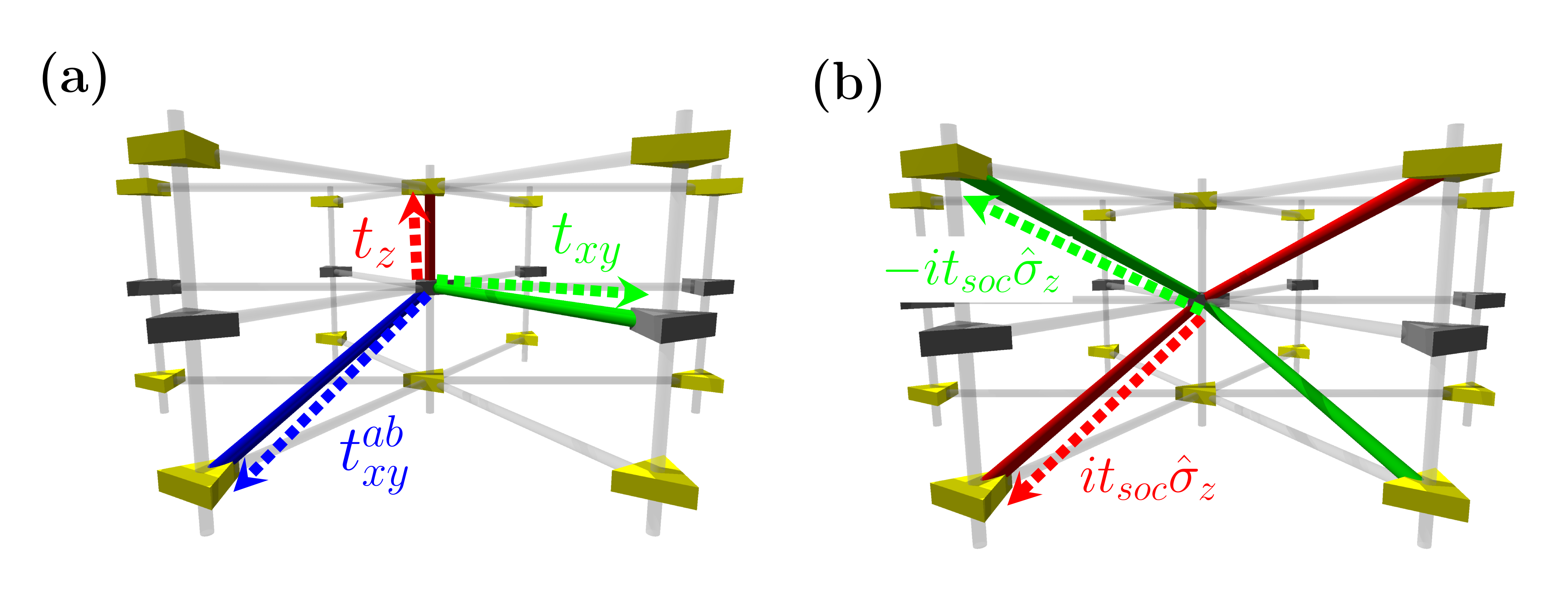}
\caption{ (color on line)
Hopping parameters of the tight-binding model for the (a) spin-independent and (b) spin-dependent hopping processes. The gray and golden triangles denote the two orbits of the model and the bonds linking them present the corresponding hopping parameters.
\label{fig:hopping} }
\end{figure}

\begin{figure}[!thb]
\centering{}
\includegraphics[width=0.5\textwidth]{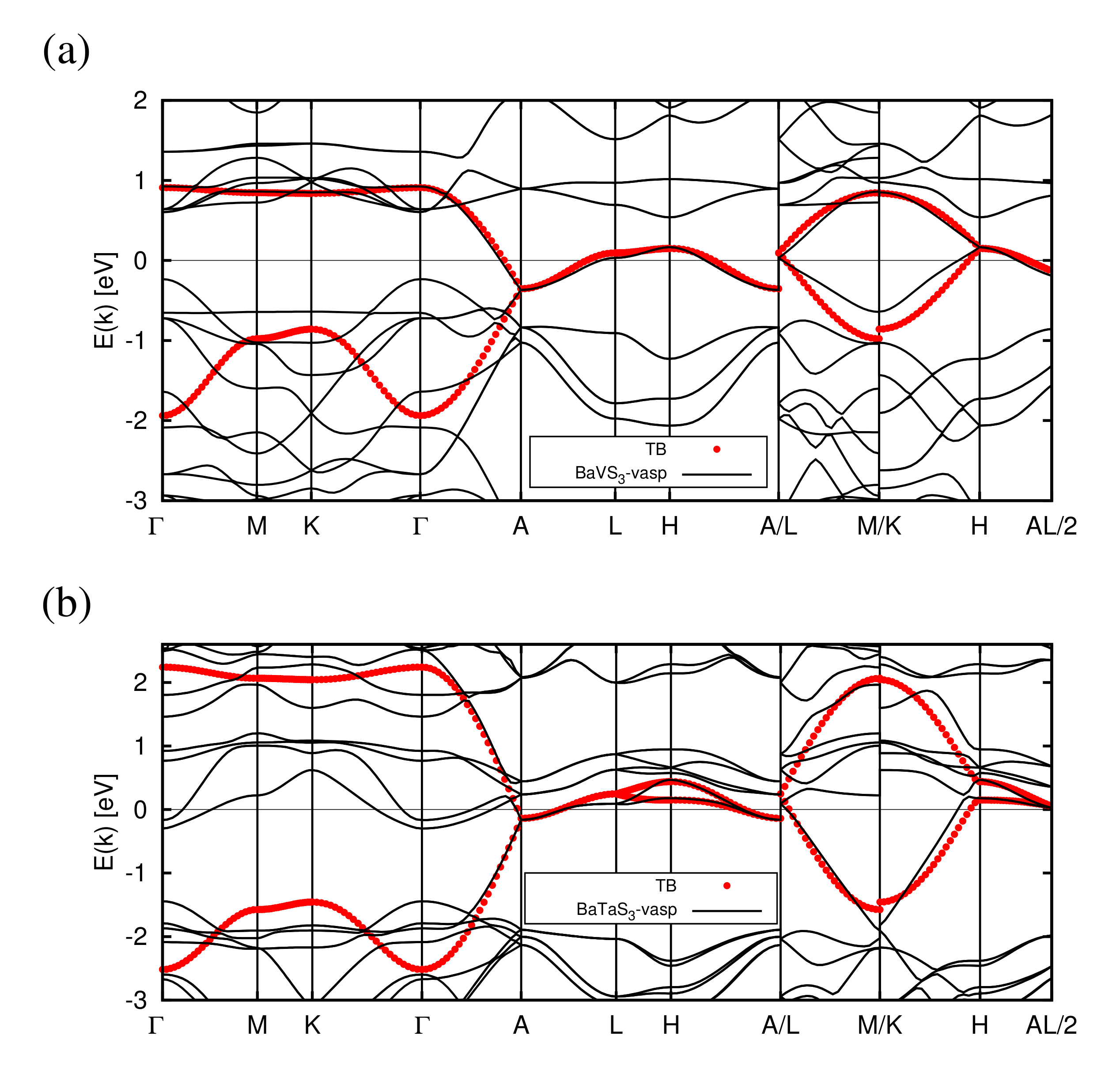}
\caption{ (Color online)
Fitting of the ab initio band structure by the ting-binding model for spinless case of BaVS$_3$ (top) and the spinful case of BaTaS$_3$ (bottom). Fitting parameters are t$_{z}$ = -0.52 eV, t$_{xy}^{ab}$ = -0.032 eV and t$_{xy}$ = -0.056 eV for BaVS$_3$ and t$_{z}$ = -0.90 eV, t$_{xy}^{ab}$ = -0.040 eV, t$_{xy}$ = -0.055 eV and t$_{soc}$= 0.062 eV for BaTaS$_3$.
\label{fig:tbfit} }
\end{figure}
In this section we construct a tight-binding model that is fully consisted with the symmetry constraints discussed above. The active orbits are the two local $dz^2$ states centered on the Ta atoms and they provide the four local basis states when spin degree of freedom is included. Four hopping parameters are given to describe the $dz^2$ bands near the fermi surface and their definitions are presented in Fig.\ref{fig:hopping}(a) schematically. The two types of Ta atoms are denoted by triangles pointing at different orientations. The hopping process along the TaS$_3$ chains is denoted by $t_z$. In the $x-y$ plane, hopping process take places between the same Ta -sublattice is defined as $t_{xy}$  while the hopping between different sublattice in the slant direction is defined as $t_{xy}^{ab}$. There are also spin-dependent hoppings when the spin orbit coupling is taken into consideration.

With these hopping parameters we are able to construct a $4\times4$ Hamiltonian $H_{TB}(\mathbf{k})$ as,
\begin{eqnarray}
H_{TB}(\mathbf{k})=\left[\begin{array}{cccc}
g(\mathbf{k}) & u(\mathbf{k})+v(\mathbf{k})& 0 & 0 \\
    & g(\mathbf{k}) & 0 & 0 \\
    &  & g(\mathbf{k}) &  u^*(\mathbf{k}) +v^*(\mathbf{k})    \\
 \dag &   &  & g(\mathbf{k})   \\
\end{array}\right], \nonumber \\
\label {eq:H_tb}
\end{eqnarray}
where
\begin{eqnarray}
g(\mathbf{k})&=&2t_{xy}\left[   \cos\mathbf{k}\cdot\mathbf{a}+\cos\mathbf{k}\cdot\mathbf{b}
+\cos\mathbf{k} \cdot (\mathbf{a}+\mathbf{b})\right]  \nonumber\\
&=&2t_{xy}\left[2\cos\frac{k_xa_0}{2}\cos\frac{\sqrt{3}k_ya_0}{2}+\cos{k_xa_0}\right]  \nonumber\\
v(\mathbf{k})&=&2it_{soc}\sin\frac{\mathbf{k}_z\mathbf{c}}{2}[   \sin\mathbf{k}\cdot\mathbf{a}+\sin\mathbf{k}\cdot\mathbf{b}
-\sin\mathbf{k} \cdot (\mathbf{a}+\mathbf{b})]  \nonumber\\
&=&2it_{soc}\sin\frac{k_zc}{2}\left[2\sin\frac{k_xa_0}{2}\cos\frac{\sqrt{3}k_ya_0}{2}-\sin{k_xa_0}\right]  \nonumber\\
u(\mathbf{k})&=&t^{ab}_{xy}\cos\frac{\mathbf{k}_z\cdot\mathbf{c}}{2}\left[   \cos\mathbf{k}\cdot\mathbf{a}+\cos\mathbf{k}\cdot\mathbf{b}
+\cos\mathbf{k} \cdot (\mathbf{a}+\mathbf{b})\right] \nonumber\\
&+&2t_z\cos\frac{\mathbf{k}_z\cdot\mathbf{c}}{2} \nonumber\\
&=&2t^{ab}_{xy}\cos\frac{k_zc}{2}\left[2\cos\frac{k_xa_0}{2}\cos\frac{\sqrt{3}k_ya_0}{2}+\cos{k_xa_0}\right] \nonumber \\
&+& 2t_z\cos\frac{k_zc}{2} \nonumber \\
\label{eq:diag_6}
\end{eqnarray} with $\hat{a}=(\frac{1}{2},-\frac{\sqrt{3}}{2},0)a_0$
and $\hat{b}=(\frac{1}{2},\frac{\sqrt{3}}{2},0)a_0$. Here $a_0$ is the latticelattice  constant in the $x-y$ plane. It is easily verified that at high symmetry paths $k_x=0$ and $k_x$ = $\pm{\sqrt{3}}k_y$ on the $k_z=\pi/c$ plane, $u(\mathbf{k})=v(\mathbf{k})=0$, indicating that there exist three node-lines.

We also fit the \textit{ab initio} band structure shown in the main text with this TB-model and the results are shown in Fig. \ref{fig:tbfit}. With suitable parameters, the \textit{ab initio} band structures are fitted qualitatively well. As the \textit{ab initio} band structures are strongly hybridized to the other valence bands, we are more focused on the bands along the path $\mathrm{A-L-H-A}$.

\bibliographystyle{naturemag}
\bibliography{refs}
\end{document}